# Twisted Lattice Nanocavity Based on Mode Locking in Momentum Space

Ren-Min Ma[1,2,3,4*], Hong-Yi Luan[1], Zi-Wei Zhao[1], Wen-Zhi Mao[1], Shao-Lei Wang[1], Yun-Hao Ouyang[1], Zeng-Kai Shao[1]

[1] State Key Laboratory for Mesoscopic Physics and Frontiers Science Center for Nano-optoelectronics, School of Physics, Peking University, China
[2] Collaborative Innovation Center of Quantum Matter, Beijing, China
[3] Peking University Yangtze Delta Institute of Optoelectronics, Nantong, Jiangsu, China
[4] National Biomedical Imaging Center, Peking University, China
* Correspondence to: renminma@pku.edu.cn

**Abstract: Simultaneous localization of light to extreme spatial and spectral scales is of high importance for testing fundamental physics and various applications. However, there is a long-standing trade-off between localizing light field in space and in frequency. Here we discover a new class of twisted lattice nanocavities based on mode locking in momentum space. The twisted lattice nanocavity hosts a strongly localized light field in a 0.048 $\lambda^3$ mode volume with a quality factor exceeding $2.9\times10^{11}$ (~250 μs photon lifetime), which presents a record high figure of merit of light localization among all reported optical cavities. Based on the discovery, we have demonstrated silicon based twisted lattice nanocavities with quality factor over 1 million. Our result provides a powerful platform to study light-matter interaction in extreme condition for tests of fundamental physics and applications in nanolasing, ultrasensing, nonlinear optics, optomechanics and quantum-optical devices.**

**Introduction**

From Planck's law on blackbody radiation to Fermi Golden rule and Purcell effect, the notion that a radiation process depends not only on the intrinsic properties of an emitter but also on its surrounding environment lays the foundation for the understanding and developing of light-matter interaction related fields [1-6]. However, light matter interaction is in general a weak process due to the mismatched wavelength of photons and electrons. In decades, artificial microstructures represented by metamaterials have been developed to enlarge and manipulate photon density of states to enhance light matter interaction [3,7-19]. In terms of that, the ratio of quality factor over mode volume (Q/V) represents a key figure of merit, because it characterizes how strong a light field can be confined in both spatial and spectral scales.

Because the applicable finite potential well in photonics, there is a long-standing trade-off between confining light field in space and in frequency. Localization mechanisms including total internal reflection [20-27], photonic bandgap [28-37], plasmonic resonance [38-43] and bound states in the continuum (BICs) [44-48] have been developed for decades to design optical cavities with high Q/V for high-performance lasers, nonlinear optics, optomechanics and quantum-optical devices and so on. However, the trade-off limits the highest figure of merit of Q/V available. Dielectric whispering-gallery-mode microcavities based on total internal reflection can achieve quality factor over a billion, but with a mode volume orders of magnitude larger than $\lambda^3$ ($\lambda$: free space wavelength) [20,22,27]. Plasmonic nanocavity employing atomistic protrusion on a host nanoparticle can achieve mode volume as small as 1 nm$^3$, but with a limited quality factor around 10 [43]. Photonic crystal nanocavities can achieve near diffraction-limited mode volume, but to achieve high quality factor needs complicated design for a full band gap and judiciously tuned wavefunction [28-37]. For instance, deep learning has been employed to optimize the quality factor of photonic crystal nanocavities, where a Q factor of $1.58\times10^9$ was obtained after optimizing the positions of 50 holes over $\sim10^6$ iterations [35]. BICs can be used to localize light field in one out-of-plane dimension with infinite quality factor but at the cost of a fully delocalized field in the other two lateral dimensions. With lateral confinement to have a finite V, the highest Q/V realized in BICs cavities is about $2.5\times10^6$ $\lambda^{-3}$, where the quality factor and mode volume are $1.09\times10^6$ and $0.43$ $\lambda^3$ respectively [48].

Recently, flatband induced wavefunction localization in moiré superlattices has drawn great attention in electronic [49-51], photonic [52-57] and phononic systems [58-63]. Compared to conventional laser cavities where discontinuity of material property or disorder is required of light field localization, flatband induced field localization can be realized in periodic moiré superlattices. However, notwithstanding the fast development of work on photonic moiré cavities, the highest Q/V achieved is yet many orders of magnitude lower than conventional laser cavities.

Here, we reveal a non-flatband effect field localization mechanism of mode locking in momentum space in artificial twisted lattice system, and demonstrate a new class of twisted lattice nanocavities with Q/V over $6\times10^{12}$ $\lambda^{-3}$, which is more than one order of magnitude higher than all reported optical cavities. We find that the twisted lattice supplies an adiabatic potential for strongly localizing light field in a deep subwavelength scale with ever increasing quality factor with the decreasing of the twisted angle. In stark contrast to the well-studied flatbands induced mode localization, twisted lattice nanocavities have much smaller footprint and are free from periodicity requirement - the twisted angle between the two sets of lattices can be arbitrary, which greatly simplifies fabrication procedure and provides a powerful platform for studying light-matter interaction in extreme condition and applications. To demonstrate the applications of twisted lattice nanocavities, we have constructed silicon based twisted lattice nanocavities silicon on insulator (SOI) substrates, where the experimentally demonstrated quality factor is over 1 million.

**Strong field localization based on mode locking in momentum space**

We use photonic graphene lattices to construct the twisted lattice nanocavities (Fig. 1a). In the structure, two sets of photonic crystals with a twist are introduced to the same dielectric membrane [54] to simplify the fabrication procedure and to enhance the coupling strength between Bloch modes of the two sets of the photonic crystals. For those twisted angles that periodic moiré superlattices can be constructed, we use a single moiré unit cell to construct the nanocavities. For those twisted angles that periodic moiré superlattices cannot be constructed, we use a quasi-single moiré unit cell. For the constructed nanocavities, a smaller twisted angle corresponds to a larger physical cavity size.

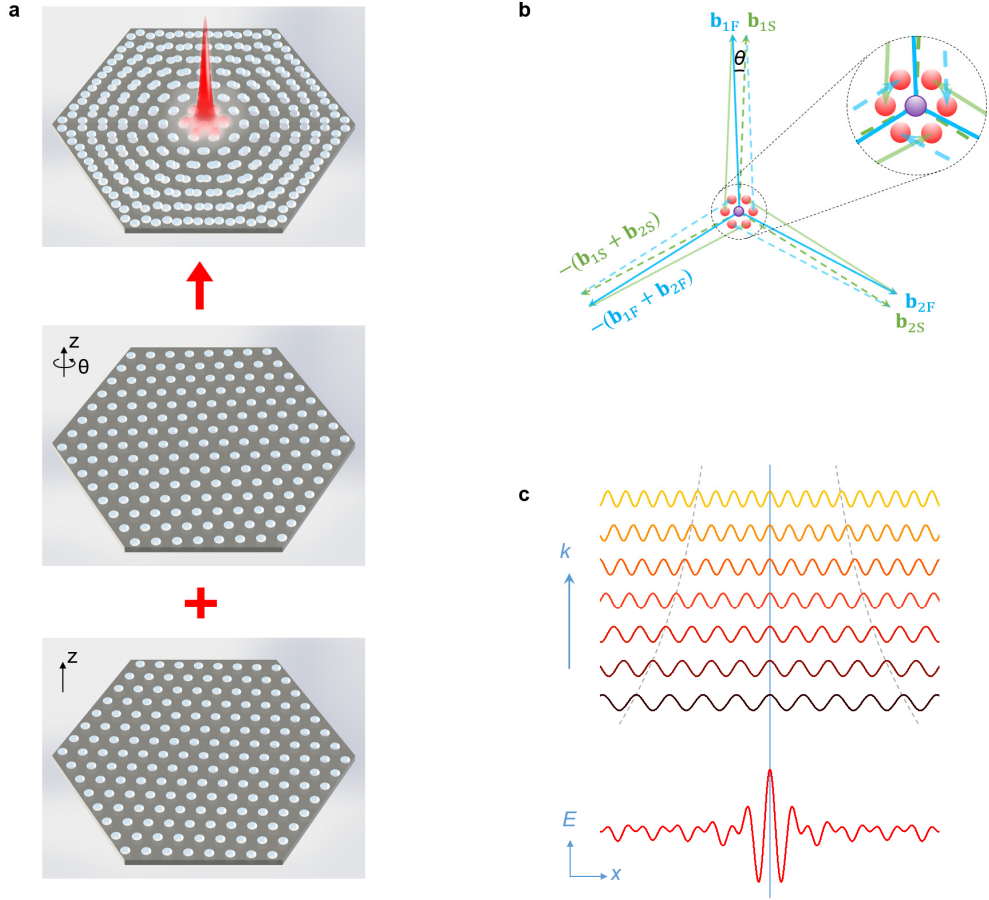

**FIG. 1. Field localization of twisted lattice nanocavities by mode locking in momentum space. (a)** Schematic of a twisted lattice nanocavity (Top) which is constructed by two sets of truncated photonic crystals with a twist (Bottom two panels). **(b)** Schematic of Bloch modes coupling induced by reciprocal lattice vectors of the two sets of twisted photonic crystals, where the Bloch mode indicated by purple dot is coupled to six other Bloch modes indicated by red dots via $\pm(\mathbf{G_F}-\mathbf{G_S})$. $\mathbf{G_F}$: reciprocal lattice vectors of the first set of photonic crystal. $\mathbf{G_S}$: reciprocal lattice vectors of the second set of photonic crystal. Three blue arrows indicate $\mathbf{G_F}$ of $\mathbf{b_{1F}}$, $\mathbf{b_{2F}}$ and $-(\mathbf{b_{1F}}+\mathbf{b_{2F}})$, and three dashed green arrows indicate $\mathbf{G_S}$ of $\mathbf{b_{1S}}$, $\mathbf{b_{2S}}$ and $-(\mathbf{b_{1S}}+\mathbf{b_{2S}})$, where $\mathbf{b_{1F}}$, $\mathbf{b_{2F}}$ are two basis vectors of $\mathbf{G_F}$, and $\mathbf{b_{1S}}$, $\mathbf{b_{2S}}$ are two basis vectors of $\mathbf{G_S}$. **(c)** Schematic of the coupling induced mode locking in momentum space, which results in Bloch modes localization in real space. $E$, $x$, and $k$ are electric field, position and wavevector respectively.

The strong field localization in twisted lattice nanocavities originates from Bloch mode coupling induced by reciprocal lattice vectors of the two sets of twisted photonic crystals (Fig. 1b). In a single set of photonic crystal, Bloch modes are delocalized waves spanning the whole area of the photonic crystal, and only these modes differing in a reciprocal lattice vector (denoted as $\mathbf{G_F}$) can couple to each other. After introducing the second set of photonic crystal, a Bloch mode can couple to other Bloch modes when their momenta differ in $\pm(\mathbf{G_F}-\mathbf{G_S})$, where $\mathbf{G_S}$ is denoted as reciprocal lattice vectors of the second set of photonic crystal. The coupling induces mode locking in momentum space, which results in localization of originally delocalized Bloch modes in real space (Fig. 1c).

We can see that the localization mechanism of twisted lattice nanocavities does not put any constraints on the twisted angle – one can continuously change the twist angle to obtain a localized light field. As a comparison, to form a moiré superlattice, the twisting has to be made in a certain set of discrete angles, where $\pm(\mathbf{G_F}-\mathbf{G_S})$ becomes moiré reciprocal lattice vectors corresponding to moiré periodicity.

A photonic graphene lattice can be viewed as a triangular lattice consisting of hexagonal unit cell with 6 sites. We use 2 degenerate dipole modes of $|p_x\rangle$ and $|p_y\rangle$ and 2 degenerate quadrupole modes of $|d_{x^2-y^2}\rangle$ and $|d_{xy}\rangle$ of one unit cell of 6 sites as Wannier functions to construct Bloch modes in a single set of photonic graphene lattice. Due to the interlayer coupling by $\pm(\mathbf{G_F}-\mathbf{G_S})$, these four modes will become localized modes in the center of twisted lattice nanocavities. Throughout the work, we focus on the localized dipole modes with smaller mode volume in twisted lattice nanocavities.

**Ultrahigh figure of merit of quality factor over mode volume**

Fig. 2a shows field distribution patterns of a twisted lattice nanocavity at 4.41° obtained by three-dimensional full wave simulation. We can see that the dipole mode is strongly localized in the center of the nanocavity in all three dimensions. Fig. 2b and the blue curve of Fig. 2c show the mode profile and the mode volume of the dipole mode at varied twisted angles where all other parameters of the two sets of graphene photonic crystals are fixed. Clearly, the mode volume of the localized dipole mode almost does not change with the twisted angle.

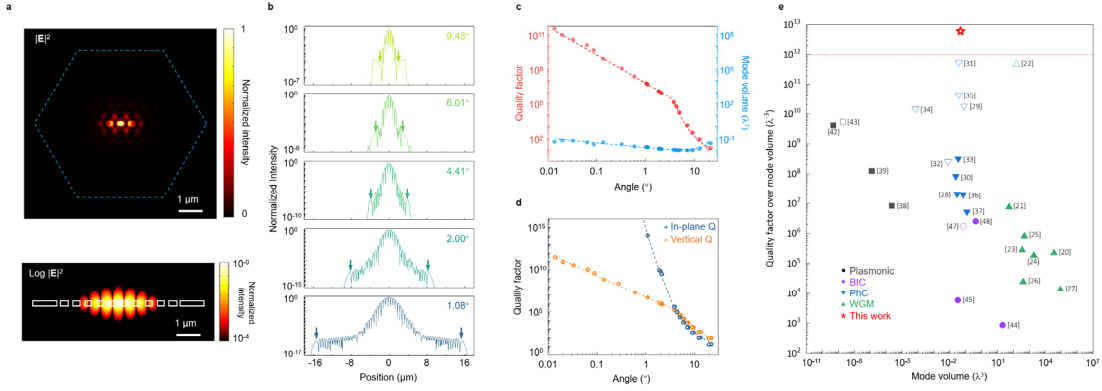

**FIG. 2. Ultrahigh figure of merit of field localization of twisted lattice nanocavities. (a)** Electric field intensity $|E|^2$ distribution of the localized mode induced by mode locking in momentum space. Blue dashed hexagon indicates the cavity boundary. Bottom: Electric field intensity distribution in log scale in a cross-section of the cavity. White boxes: contour of the cavity structure. **(b)** Intensity distributions of the localized dipole modes in log scale at five twisted angles. The mode keeps being strongly localized with the decreasing of the twisted angle. Arrows indicate cavity boundaries. **(c)** The scaling laws of quality factor (Q) and mode volume (V) vs. twisted angle obtained by three-dimensional full wave simulation where all the parameters of the two sets of graphene photonic crystals are fixed. Dots and circles represent angles that periodic moiré superlattices can and cannot be constructed respectively. **(d)** Scaling laws of in-plane and vertical quality factor of twisted lattice nanocavities. Dots: data. Lines: fitting. **(e)** Comparison of quality over mode volume (Q/V) with other representative optical cavities of whispering-gallery-mode (WGM) microcavities, photonic crystal (PhC) nanocavities, plasmonic nanocavities, bound-state-in-the-continuum (BIC) microcavities. Hollow: quality factor obtained by simulation. Solid: quality factor obtained by experiment.

Remarkably, the quality factor of the localized dipole mode continuously increases with the decreasing of the twisted angle (red curve of Fig. 2c). As shown in figure, the scaling of quality factor over twisted angle has a kink around 4.41º, which originates from the transition from in-plane scattering loss dominant region to out-of-plane scattering loss dominant region (Fig. 2d). A larger angle gives a smaller cavity size, which results in a comparably stronger field at cavity boundaries that can be coupled to free propagating modes of the membrane, leading to a larger in-plane scattering loss. At smaller twisted angles, cavity size becomes larger, therefore such in-plane scattering loss substantially decreases. Around the twisted angle of 4.41º, out-of-plane scattering loss becomes dominant.

Interestingly, a smaller twisted angle will also suppress out-of-plane scattering loss to free space. With the decreasing of the twisted angle, the arrangement of nanoholes from the center to the boundaries of a cavity changes more slowly (SEM images in Fig. 3a-e). However, the intensity distribution of the wavefunction of the localized dipole modes has almost no change - it keeps being tightly localized in the center of the nanocavities with the decreased angle. Therefore, at a smaller angle, the wavefunction of a localized dipole mode experiences a more effective adiabatic change in the arrangement of nanoholes towards cavity boundaries, leading to a lower scattering loss to free space and a higher vertical Q. In the out-of-plane scattering loss dominant region, we can use a truncated cavity with a relatively smaller physical size for easier fabrication while maintaining a high quality factor.

The quality factor reaches to $2.9 \times 10^{11}$ at $0.0138°$, with a corresponding mode volume of $0.048\,\lambda^3$, which results in a Q/V of $\sim 6 \times 10^{12}\,\lambda^{-3}$ which is more than one order of magnitude higher than all reported optical cavities to our knowledge (Fig. 2e). The highest Q/V presented in Fig. 2 is only limited by our computing power in three-dimensional full wave simulation. The twisted lattice nanocavities can achieve ever increasing Q/V by just simply making the twisting angle smaller, which provides a powerful platform to pursuit ever increasing field enhancement in a nanocavity.

**Silicon based twisted lattice nanocavities**

Experimentally, we fabricate twisted lattice nanocavities in silicon membrane from SOI substrates. Fig. 3a-e show the scanning electron microscopy (SEM) images of twisted lattice nanocavities at five twisted angles from $0.596°$ to $5.08°$. The red and yellow circles indicate two lines of nanoholes from the twisted two graphene lattices. A tunable continuous wave laser is used to excite the localized dipole mode in the fabricated nanocavities. While all these cavities hold a localized mode in the center area that can be excited under resonance condition, the quality factor of the mode increases with the decreasing angle because the more adiabatic potential as discussed above.

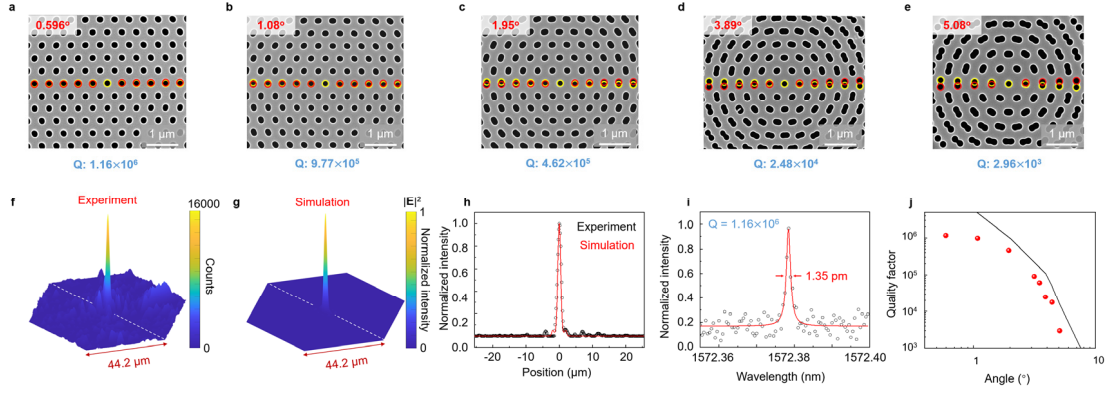

**FIG. 3. Silicon based twisted lattice nanocavity. (a-e)** SEM images of silicon based twisted lattice nanocavities at five twisted angles from 0.596° to 5.08°. Red and yellow circles guide the eye to the twisting of the lattices. **(f-g)** Experimentally excited (f) and full wave simulated (g) localized dipole mode of the cavity at twisted angle of 0.596°. **(h)** Intensity distributions of the localized mode along the white dashed lines in (f) and (g). **(i)** Experimentally measured scattering spectrum of the nanocavity at twisted angle of 0.596°. The resonance wavelength and linewidth of the mode are 1572.38 nm and 1.35 pm respectively, which gives a quality factor over 1 million. **(j)** Experimentally obtained quality factor for cavities at 8 different twisted angles. Line: simulation result.

Fig. 3f-g show the experimentally excited and full wave simulated patterns of the localized dipole mode in the cavity at the twisted angle of 0.596°, which matches well with each other (Fig. 3h). In the two sets of twisted photonic graphene lattices of the cavity, the diameter of the nanoholes is 210 nm, and the lattice constant is 460 nm. The thickness of the silicon membrane is 220 nm. The resonance wavelength and linewidth of the mode are 1572.38 nm and 1.35 pm respectively (Fig. 3i), which yields a quality factor of ~$1.16 \times 10^6$, a value among the highest quality factors obtained experimentally for nanocavities.

Fig. 3j shows the experimentally obtained quality factor for cavities at 8 different twisted angles from 0.596° to 5.08°. The obtained quality factors for the devices at 0.596° and 1.08º are close to the resolution limit of our measurement system, while the obtained quality factors for the devices at larger twisted angles are close to the simulated values. The passive silicon based twisted lattice nanocavities can be employed to study field enhancement related physics and devices in classical and quantum regimes [3-6].

**Conclusion**

We demonstrate a new class of twisted lattice nanocavities based on mode locking in momentum space which presents a record high figure of merit of light localization among all reported optical cavities. In stark to the well-studied flatband induced field localization, the twisted lattice nanocavities is free from periodicity requirement which greatly simplifies fabrication procedure. The quality factor of the nanocavity increases continuously with the decreasing of the twisted angle without fine-tuning in any other structure parameters. The highest quality factor obtained by three-dimensional full wave simulation exceeds 200 billion with a mode volume of ~0.048 $\lambda^3$. We have further constructed silicon based twisted lattice nanocavities and III-V based twisted lattice nanolasers based on the design. The measured quality factor for the silicon based twisted lattice nanocavities is over 1 million. The demonstrated twisted lattice nanocavities provides a powerful platform to study light-matter interaction for tests of fundamental physics, and promises new functional devices with unprecedented performance from classical to quantum regime.

**Acknowledgements**

This work is supported by the National Key R&D Program of China (grant no. 2018YFA0704401), the Beijing Natural Science Foundation (grant no. Z180011), and the National Natural Science Foundation of China (grant nos. 91950115, 11774014, 61521004 and 62175003).